\documentclass[aip, amsmath,amssymb, preprint]{revtex4-1}

\usepackage{CJK}

\usepackage[final]{graphicx}
\usepackage{hyperref}
\usepackage{color}
\usepackage{enumerate}

\usepackage{xr}

\usepackage{dcolumn}
\usepackage{bm}

\usepackage[utf8]{inputenc}
\usepackage[T1]{fontenc}
\usepackage{mathptmx}
\usepackage{etoolbox}
\usepackage{hyperref}
\usepackage{xcolor}
\usepackage{soul}

\makeatletter
\def\@email#1#2{%
 \endgroup
 \patchcmd{\titleblock@produce}
  {\frontmatter@RRAPformat}
  {\frontmatter@RRAPformat{\produce@RRAP{*#1\href{mailto:#2}{#2}}}\frontmatter@RRAPformat}
  {}{}
}%
\makeatother

\begin{document}
\begin{CJK*}{GB}{} 

\preprint{STS/Neutron-Optics}

\title[]{General Guide Concepts for Compact, High-Brilliance Neutron Moderators}

\author{Yaohua Liu*}
\email[The author to whom correspondence may be addressed:~]{liuyh@ornl.gov}
\affiliation{Second Target Station, Oak Ridge National Laboratory, Oak Ridge, Tennessee 37831, USA}%
\date{\today}

\begin{abstract}
The trend in neutron sciences is toward integrating compact, high-brightness moderators into new or upgraded facilities. Transporting neutrons from the source to the sample position with a phase-space distribution tailored to specific requirements is crucial to leverage high source brilliance. We have investigated four guide concepts using Monte Carlo ray tracing simulations: Montel beamline with nested Kirkpatrick-Baez mirrors, curved-tapered beamline with a bender and straight sections, straight-elliptical beamline, and curved-elliptical beamline. The straight-elliptical (curved-elliptical) beamline features two half-ellipse guides connected by a straight (non-straight) guide section. The neutron transport efficiency and phase space homogeneity have been quantitatively compared. Our results show that the straight-elliptical beamline performs best because of few neutron bounces on the guide surface with small reflection angles, minimizing flux loss. The Montel beamline provides the best spatial confinement of neutrons within the desired region; however, there is a high thermal-neutron loss due to large reflection angles. The curved-tapered beamline suffers from significant flux loss due to high bounces, and it shows a non-uniform angular distribution related to broad ranges of bounces and reflection angles. The non-straight guide section of the curved-elliptical beamline increases phase space inhomogeneity, leading to a spatially non-uniform beam profile. The results apply to general neutron instruments that require transporting thermal and cold neutrons from a compact, high-brilliance moderator to the sample location with a moderate phase-space volume.

\end{abstract}
\maketitle
\end{CJK*}

\section{\label{sec:intro}Introduction} 
Globally, countries are investing in cutting-edge, high-power accelerator-based neutron sources, including the Spallation Neutron Source (SNS)~\cite{mason2006spallation}, the European Spallation Source (ESS)~\cite{garoby2017european}, the Japan Proton Accelerator Research Complex (J-PARC)~\cite{takada2017materials}, and the China Spallation Neutron Source (CSNS)~\cite{wei2009china}. These facilities represent the pinnacle of research infrastructure, utilizing uncharged neutrons with unique properties to probe the structure and dynamics of materials at the atomic and molecular levels and beyond, often revealing insights unattainable through other techniques~\cite{carpenter2015elements}. The unparalleled capability of neutron scattering fuels advancements in diverse fields like medicine, energy, and electronics, driving scientific innovation with profound implications for technological progress and economic growth. 

The landscape of neutron sciences is increasingly defined by using compact, high-brightness moderators in the latest new and upgraded facilities, including ESS and the Second Target Station (STS) at SNS, because recent studies have found that smaller moderators yield higher brilliance for spallation sources ~\cite{zhao2013optimizing, gallmeier2016conceptual, zanini2019design}. To maximize the impact of the high-brilliance sources, neutron guides are instrumental in delivering neutron beams for experiments, influencing the brilliance, homogeneity, and spectrum. Neutron guides are made of unique metamaterials called supermirrors that reflect neutrons when they hit the surface at shallow angles, bouncing them back and keeping them within the guide. Simple neutron guides often have a constant and rectangular cross-section. Supermirrors have imperfect reflectivity, and modern guides often have non-trivial geometries to either utilize or mitigate the effects of this imperfection to achieve desired transport performance~\cite{mezei2000neutron, schanzer2004advanced}. A curved guide can avoid the direct line of sight of the moderator, which behaves as a low-energy-neutron-pass filter that can remove unwanted high-energy neutrons from the beam~\cite{maier1963use}. Alternatively, a ballistic guide can transport neutrons over an extended distance with few bounces that minimize beam loss~\cite{mezei1997raison}. With the advancement of neutron ray tracing software, it becomes common to integrate Monte Carlo simulations into the guide design that optimizes sophisticated geometries to maximize the neutron flux at the sample location with the desired wavelength range, beam size and divergence, and profile homogeneity. On the other hand, advancements in manufacturing make it feasible to make guides with non-trivial geometries. This synergy of computational and manufacturing capabilities becomes crucial for optimizing neutron transport to leverage high-brilliance sources that underpin scientific breakthroughs. 

We have compared four general guide concepts using a compact, high-brilliance moderator, exploring the effects of guide geometries and imperfect mirror reflection. This work focuses on the STS single-crystal neutron diffractometer concept Pioneer, which requires a moderate phase-space volume at the sample position~\cite{liu2022Pioneer}. The STS represents a significant advancement for neutron sources, joining the existing sources of the SNS First Target Station and the High Flux Isotope Reactor at the Oak Ridge National Laboratory. It fills crucial capability gaps by providing a high-intensity, cold neutron source with a peak brightness exceeding all existing and planned neutron sources~\cite{adams2020first}. This enhancement will empower STS instruments to deliver transformative performance. Pioneer will be a diffractometer for studying small single crystals and ultra-thin epitaxial films in versatile sample environments~\cite{liu2022Pioneer}. It will use neutron guides to achieve three objectives: (1) high flux on samples to enhance signals, (2) low flux outside the phase-space region of interest (ROI) to match the resolution needs and minimize background scattering, and (3) high spatial and angular homogeneity within ROI to maximize experimental accuracy. These objectives are universal in achieving high-quality data for many neutron scattering instruments, although certain techniques are less sensitive to phase space inhomogeneity. 

The four neutron guide concepts include  (1) a Montel beamline with a pair of nested Kirkpatrick-Baez mirrors, (2) a curved-tapered beamline with a bender, (3) a straight-elliptical beamline with a T$_{0}$ chopper, and (4) a curved-elliptical beamline. The straight-elliptical (curved-elliptical) beamline employs two half-ellipse guides connected by a linearly-tapered, straight (non-straight) guide section. The guide geometries are numerically optimized with Monte Carlo ray tracing simulations. The neutron transport efficiencies are quantified using the brilliance transfer (BT) concept~\cite{andersen2018optimization}. We have analyzed the details of supermirror reflections and the phase space homogeneity to gain insights into the distinct performance of individual guides.   

The straight-elliptical beamline with two half-ellipse guides has the highest BT and a uniform phase space. The Montel beamline performs best in confining neutrons within the spatial ROI but has the lowest thermal-neutron transport efficiency. The curved-tapered beamline has the lowest cold-beam transport efficiency, and the curved-elliptical beamline experiences high spatial inhomogeneity. The curved-tapered and curved-elliptical beamlines also deliver more neutrons outside the desired spatial ROI. Our post-optimization analysis shows that the superior transport efficiency of the straight-elliptical beamline is related to low bounces on the guide surface and small reflection angles, minimizing the flux loss. The non-straight guide section used by the curved-elliptical beamline increases bounces and reflection angles and leads to more beam loss and a spatially non-uniform beam profile. The excessive thermal-neutron loss using Montel mirrors is due to high reflection angles. For the curved-tapered beamline, high bounces limit the  BT, and a broad distribution of bounces and reflection angles leads to a non-uniform angular distribution. Our results can be generalized to neutron scattering instruments facing a compact, high-brilliance moderator, using cold and thermal neutrons and requiring a moderate phase-space volume at the sample location.

\section{\label{sec:method} Methodology} 

\subsection{\label{sec:method_req}General Requirements} 
\begin{figure}
\includegraphics[width=0.5\textwidth]{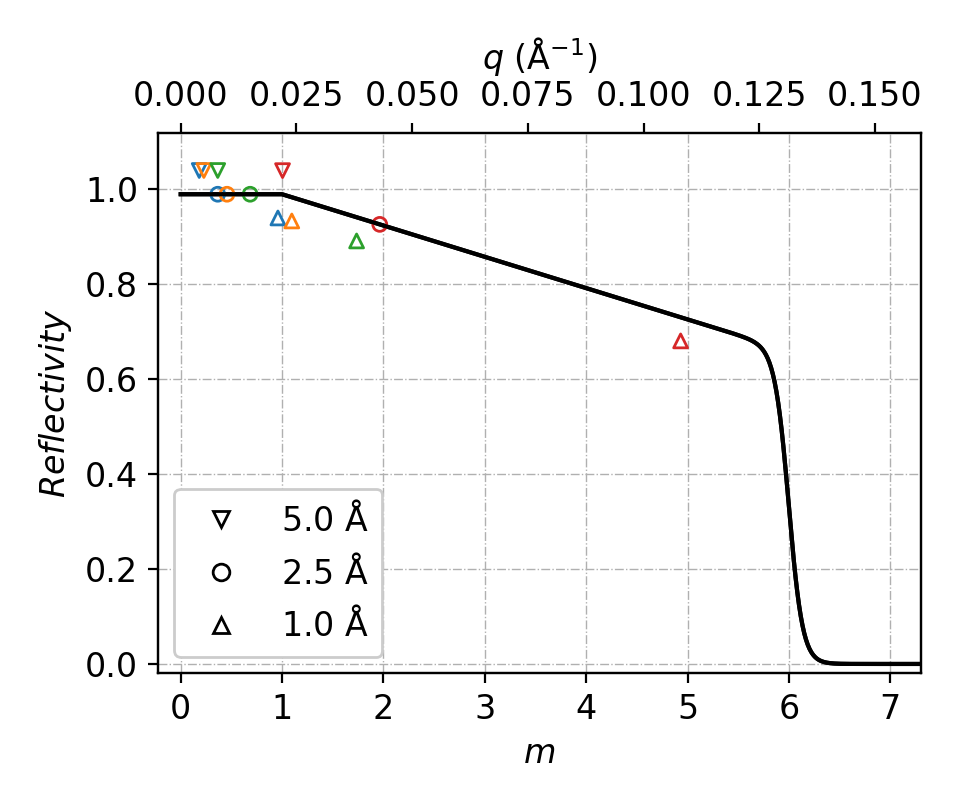}
\caption{\label{fig:Refl} The calculated reflectivity curve of a $m=6$ supermirror is used in our Monte Carlo ray tracing simulations. The symbol indicates the locations of the averaged momentum transfers for three representative wavelengths,  1.0~\AA, 2.5~\AA~and 5.0~\AA, for the four optimized guides. Different colors denote results from different guides: blue---the straight-elliptical beamline; orange---the curved-elliptical beamline; green---the curved-tapered beamline; and red---the Montel beamline.}
\end{figure}

When high-energy protons strike the target in a spallation neutron source, a flash of fast neutrons and energetic gamma rays are released through spallation~\cite{carpenter2015elements}. These high-energy radiations easily pass through the moderator with minimal attenuation. If they reach the detector area, they will interfere with experiments with a window of unusable time-of-flight. To eliminate these unwanted particles from the beam, neutron transport systems can be curved to avoid the direct line of sight from the moderator.  An alternative approach is to use a T$_{0}$ chopper, which acts like a high-speed shutter to block the initial burst of fast neutrons and gamma rays. Still, it allows the already-moderated, slower neutrons to pass through to the sample location. This work has considered four guide concepts using either one of two options. 

Neutron scattering is, in general, a flux-limited technique. The ideal neutron transport system should deliver the highest possible neutron flux to the sample location while maintaining the desired beam size, divergence, and wavelength band. The system must minimize neutron flux outside the spatial ROI to reduce background scattering from the sample environment and holder. Reducing flux outside the divergence ROI is required to meet the reciprocal-space resolution requirement. At the same time, achieving uniform spatial and angular distributions within the ROI will improve experimental accuracy. Spatial homogeneity ensures the neutron spectrum obtained from calibration samples can be applied to experiments with actual science samples, regardless of their distinct sizes or minor misalignment, which will limit systematic errors.  Angular homogeneity minimizes ambiguity in fine scattering features and facilitates pattern fitting.

\subsection{\label{sec:method_opt}Numerical Optimization}
Advancements in computational tools and neutron optics have enabled guide design with non-trivial geometries using Monte Carlo ray tracing simulations. The critical aspect of the optimization is choosing metrics. We have considered an ROI of four-dimensional (4D) phase space that includes a $5\times5$~mm$^{2}$ area and a $\pm 0.3^\circ$ beam divergence in both the horizontal and vertical directions, following the Pioneer's requirements~\cite{liu2022Pioneer}, and have used the following metric to take account into the three objectives, 
\begin{align*} 
metric~=~I_{roi}& - (w_{0} \times I_{outside-div-roi} + w_{1} \times I_{outside-spatial-roi})\\
                & - (w_{2} \times \delta I_{div-roi} + w_{3} \times \delta I_{spatial-roi}),
\end{align*} 
where $I_{roi}$ is the integrated flux within the desired 4D phase space volume; $I_{outside-div-roi}$ and $I_{outside-spatial-roi}$ are the unwanted flux outside the angular and spatial ROIs, respectively;  $\delta I_{div-roi}$ and $\delta I_{spatial-roi}$ are the standard deviations of the angular and spatial distributions, respectively, within the spatial ROI; and $w_0 - w_3$ are the hyper-parameters to control weighting factors for the unwanted flux and beam inhomogeneity. The metric was calculated by integrating across the entire spectrum of simulated neutron wavelengths. The spatial and angular degrees of freedom are decoupled to reduce the computational cost when calculating the unwanted neutrons and the phase space inhomogeneity (the $w_0$--$w_3$ terms). Neutron profiles are integrated over the spatial ROI to calculate the angular distribution and are integrated over the full divergence range to calculate the spatial distribution. For statistical analysis, the two-dimensional angular and spatial profiles were binned evenly. The typical spatial and angular bin sizes used to calculate $\delta I$ are 100-200~$\mu$m and 0.02-0.04$^\circ$, respectively.

We have used the McStas~\cite{willendrup2014mcstas}, MCViNE~\cite{lin2016mcvine} software packages for Monte Carlo ray tracing simulations and used the SciPy~\cite{2020SciPy-NMeth}, and JupyterLab~\cite{kluyver2016jupyter} for the numerical optimization and post-optimization data analysis. Two global optimization algorithms, differential evolution and particle swarm optimization, and one local optimization algorithm, Powell, were depolyed for optimization. The global optimization algorithms were used to explore a broad parameter space, and the local optimization algorithm was used for all final optimizations. Figure~\ref{fig:workflow} shows the workflow during the final optimization, which employs a double-loop structure.  The outer loop updates the hyperparameters and resets the direction vectors, while the inner loop optimizes the instrument parameters. All guide section dimensions were freely optimized. The guide system was normally constrained to start at least 2.0~m from the source and end at least 0.8~m from the nominal sample position. However, these constraints are loose since the optimized guides start much farther and end much earlier. Additional constraints for individual guide types will be mentioned below in Sec.~\ref{sec:Opt_guide}.

This project used a cluster with 256 CPU cores (base clock 2.0~GHz ) and 1~TB of memory. One typical optimization of a single guide variant used 800 to 8000 CPU hours and took thousands to tens of thousands of iterations to converge. Due to the scholastic nature of Monte Carlo Simulations, each optimization gave different results. We have optimized 60-120 variants for each guide type. The total project used about 600000 CPU hours; however, a significant portion of the time was spent investigating the relative sensitivity of parameters and the effective limits of the parameter spaces, fine-tuning the models, and developing the optimization strategy. The best variants of each guide type have different geometries but comparable performance, suggesting the results are close to an optimum. 

\subsection{\label{sec:method_MC}Monte Carlo Ray Tracing} 
Our simulations use the STS cylindrical para-hydrogen moderator source file with a uniform view of $3\times3$~cm$^2$, which has a peak flux at 2.5~\AA~\cite{adams2020first}; therefore, the neutrons near the peak of the flux distribution effectively have more weight in the metric. The Pioneer beamline concept has a source-to-sample distance of 60.0~m and requires an operational wavelength range of 1.0--6.0 \AA~\cite{liu2022Pioneer}. A 4.3-\AA~wavelength band will be selected using two disk choppers during normal operation. A broader spectrum range, typically 0.8-6.4~\AA, is used during simulations.

Gravity effects and imperfect mirror reflection have been included in our Monte Carol ray tracing simulations. Non-linear guide shapes are modeled using the McStas component $Elliptic\_guide\_gravity()$ or using straight mirror segments with $Guide\_gravity()$. Neutron supermirrors do not have 100\% specular reflectivity. Our Monte Carlo simulations have modeled this imperfect reflectivity using the following function~\cite{willendrup2014mcstas}: $R(q) = R_{0}(1 - \alpha (q - q^{Ni}_{c}))(1-tanh((q - q_{c})/w))/2$ for $q > q^{Ni}_{c}$, where $q^{Ni}_{c} = 0.0219$~\AA$^{-1}$ is the critical scattering wavevector-transfer for natural Ni. Below $q^{Ni}_{c}$, supermirrors show a high reflectivity $R_0$, close to 1. The reflectivity decreases linearly above $q^{Ni}_{c}$ until close to $q_{c} = m \times q^{Ni}_{c}$, where it transits to zero over a width of $w$, where the index $m$ specifies the working range of supermirrors. Figure~\ref{fig:Refl} plots the reflectivity curve for $m=6$ supermirrors with $R_{0}=0.99$, $\alpha=3.044$~\AA~and $w=0.003$~\AA$^{-1}$, as used in our simulations. Since Monte Carlo simulation is a stochastic process, many neutron rays must be simulated to calculate the meaningful beam homogeneity. The typical ray numbers in our simulations range between $4 \times 10^{7}$ and $4 \times 10^{8}$.   

All guides considered here have closed, rectangular cross-sections except the Montel beamline. Small gaps in guides needed for bandwidth choppers, slits, and monitors were typically ignored. However, a T$_0$ chopper will require significant space; therefore, the straight-elliptical beamline has considered a 0.45-m gap for a T$_0$ chopper in the middle of the guide. 

\subsection{\label{sec:method_ANA}Post-Optimization Analysis} 
After the guide optimization, we conducted additional Monte Carlo simulations with the optimized guide geometry, where details of the supermirror reflection events were recorded for the neutron rays reaching the sample location, including the reflection locations, the neutron velocity after reflections, the wavevector transfer during reflections, and the neutron weight factor. The weight factor used in McStas and MCViNE is a computational efficiency tool that allows the simulation of a relatively smaller number of neutron rays while still obtaining statistically accurate results about the real neutron behavior for scattering instruments~\cite{willendrup2014mcstas, lin2016mcvine}. After each supermirror reflection, the weight factor will be multiplied by the reflectivity derived from the wavevector transfer. The total bounces for each ray can be counted from the reflection events. 

We have also investigated the phase-space homogeneity. Since the flux outside the ROI is unwanted during the optimization, there is a transition region at the ROI's boundary, where the neutron density rapidly changes. Including the edges of the phase space ROI in the statistical analysis will lead to high variations. Therefore, only the central ROI is used for post-optimization statistical analysis to emphasize the intrinsic effects of different guides. The selected phase space is within $\pm2$~mm and $\pm 0.25^\circ$ along the horizontal and vertical directions. There are 8 bins per dimension and 4096 bins in the 4D phase space.  

\section{\label{sec:Opt}Optimization results}
\subsection{\label{sec:Opt_guide}Four Guide Options}
\begin{figure*}
\includegraphics[width=0.98\textwidth]{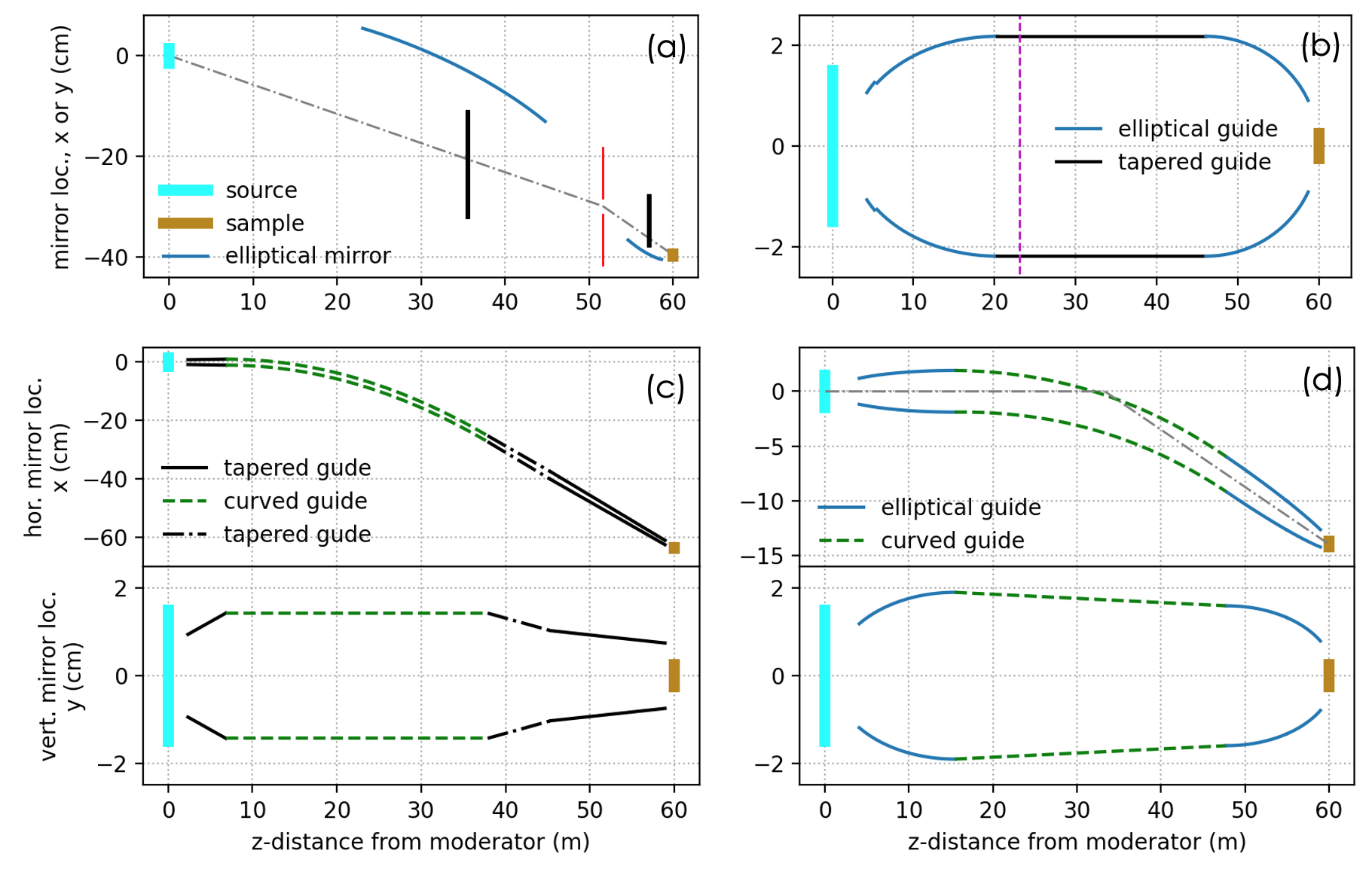}
\caption{\label{fig:Guides} Optimized guide geometries of four guide systems. (a) Montel beamline with two sets of nested Kirkpatrick-Baez mirrors.  The aqua-blue box represents the source, and the gold box represents the sample. The blue curves indicate the location of the mirror surfaces, the dot-dashed line represents the major axes of the two Montel mirrors, and the two thick black lines are built-in beam stops to block the line of sight between the moderator and the virtual source and between the virtual source and the sample, respectively. (b) A straight-elliptical beamline with a T$_{0}$ chopper, denoted by the purple dashed line, and two half-ellipse guides (blue lines) connected by a straight, linearly tapered guide section (black lines).  (c) A curved-tapered beamline with a bender (dashed green lines). There are linearly taped guides before and after the bender. (d) A curved-elliptical beamline option with two half-ellipse guides connected by a curved, linearly tapered guide section.  Both the curved-tapered and curved-elliptical beamlines are curved in the horizontal plane but are straight in the vertical direction.}
\end{figure*}

We have considered four types of guide systems, and the optimized guide geometries are shown in Fig.~\ref{fig:Guides}. The first uses two sets of Montel mirrors, as shown in Fig.~\ref{fig:Guides}a.  A Montel mirror, also known as nested Kirkpatrick-Baez (KB) mirrors, consists of two elliptical-shaped surfaces that are mutually perpendicular~\cite{ice2009nested}. Along the primary beam path, the first Montel mirror has two focal points at [0.18, 51.67]~m, and the second has them at [51.81, 59.95]~m from the moderator. A pair of slits is placed at the second focal point of the first Montel mirror to provide a virtual source. During the optimization, slit placement was constrained,  but the slit size, which determines the virtual source size, was an optimization parameter.  The horizontal and vertical mirrors have the same geometry. The first Montel set is 21.7~meters long with top and left reflective surfaces (view from the moderator), and the second is 4.0~meters long with bottom and right reflective surfaces. Since a Montel mirror has a non-closed cross-section, an in-guide beam-stop can block unwanted neutrons without breaking the mirror~\cite{liu2022Pioneer}.  However, the average beam direction is tilted away from the major axis of the Montel mirrors because there is no mirror reflection symmetry in the horizontal and vertical planes. Therefore, the mirrors' major axes are rotated to achieve a genuinely horizontal beam at the sample position. The major axes are rotated in vertical and horizontal directions, with 0.33$^\circ$ and 0.66$^\circ$, respectively, for the first and second sets of mirrors. A kink between the two major axes is introduced at the virtual source location to decrease the beam's vertical displacement and eliminate the need for lowering the floor at the sample position~\cite{liu2022Pioneer}.

Figure~\ref{fig:Guides}b shows the second system---a straight beamline with two half-ellipse guides connected by a linearly tapered, straight guide. This system requires a T$_{0}$ chopper, as denoted by the dashed purple line, because the sample is in the direct line of sight of the moderator. Locations of a $T_0$ chopper were explored before final optimizations. The results show that it performs better to have a large gap in the tapered section than in the elliptical section, which is related to the average bounces per unit length on the supermirror surfaces. Therefore,  the T$_0$ chopper's location was constrained to be within the tapered region during final optimizations. The two focal points are at [2.22, 38.56] m for the first ellipse and are at [32.18, 59.95] m for the second one along the primary beam path. Elliptical guides are superior to transport neutrons over long distances~\cite{kleno2012systematic}. Compared to the single ellipse design, the two-half-ellipse design reduces the maximum cross-section of the guide. At the same time, the long tapered section helps to minimize the divergence gaps that are unwanted features of guides only containing elliptical guide sections~\cite{weichselbaumer2015tailoring, ma2021performance}. 

The third guide system, shown by Fig.~\ref{fig:Guides}c, is a curved-straight hybrid beamline with a single-channel bender and straight guide sections~\cite{copley1995transmission, mildner1990acceptance}. The guide is straight in the vertical direction but is curved horizontally to avoid the direct line of sight from the moderator.  There are linearly taped guides before and after the bender.  The tapered section before the bender diverges in horizontal and vertical directions. There are two tapered sections after the bender with different taper angles, which offers flexibility in optimizing the guide geometry. The first section, closer to the curved guide, diverges horizontally while converges vertically. The second section, closer to the sample, converges in horizontal and vertical directions. The bender has a radius of curvature of 1834~m, a total length of 31.0~m, and a total deflection angle of 0.97$^\circ$. The bender starts at 6.84~m with a guide opening width of 2.05~cm. The total length of the bender is 31.0~m, far above the required value (17.4~m) to lose the direct line of sight, which is a good practice to reduce background~\cite{zendler2015generic}. 

The curved-elliptical beamline is modified from the straight-elliptical beamline, as shown in Fig~\ref{fig:Guides}d. In the horizontal plane, a curved, linearly tapered guide section connects the two half-ellipses, with a kinked angle of 0.3$^\circ$ between the major axes of the two half-ellipses. This curved section avoids the direct line-of-sight between the sample and the source, eliminating the need for a T$_0$ chopper. The two focal points for the first ellipse are at [0.91, 29.97] m from the moderator; and for the second one, they are at [1.77, 27.36] m from the nominal kinked point at 33.3 m. A quadratic function describes the displacement of the curved guide section in the horizontal plane. The curved guide's cross-section matches the elliptical guides at the joint points. Furthermore, the slopes of the curved section match the elliptical guides in the horizontal plane, which improves the angular distribution homogeneity.

\subsection{Brilliance Transfer}
\begin{figure}
\includegraphics[width=0.5\textwidth]{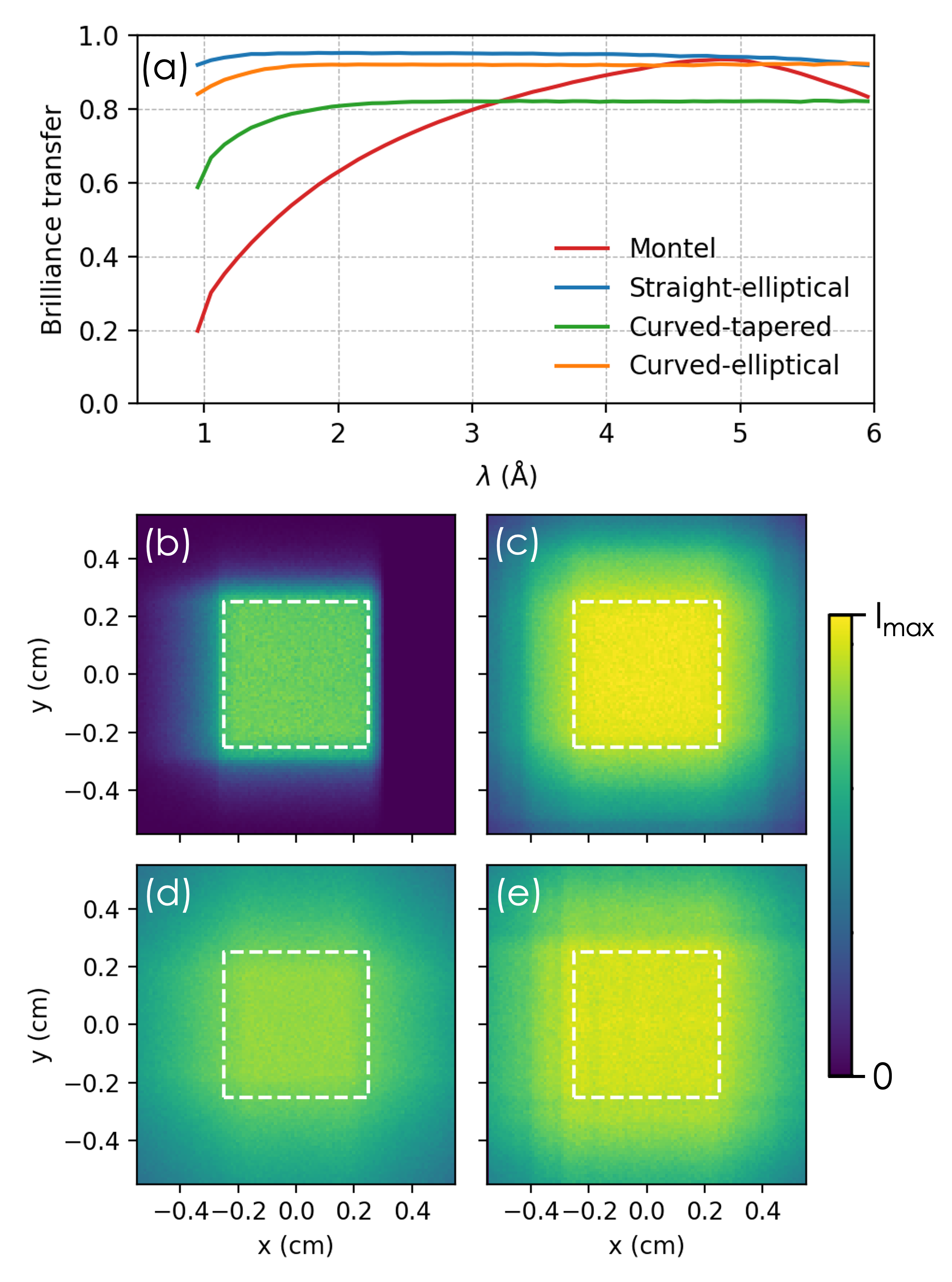}
\caption{\label{fig:BT}  (a) Wavelength-resolved brilliance transfer. (b-e) The spatial distribution of neutrons with divergences between $-0.3^{\circ}$ and $+0.3^{\circ}$ and wavelengths between 1.0~\AA~and 6.0~\AA. The bin size is $100\times100~\mu$m$^2$. (b)-(e) are the Montel, the straight-elliptical, the curved-tapered, and the curved-elliptical beamline, respectively. The white dashed boxes highlight the spatial ROIs.}
\end{figure}

Figure~\ref{fig:BT} shows the brilliance transfer and spatial distribution within the ROI of $5\times5$~mm$^{2}$ with a $\pm$0.3$^{\circ}$ beam divergence in both the horizontal and vertical directions. Neutron source brilliance, also known as neutron brightness, refers to the density of neutrons emitted from a neutron source per unit area, per unit solid angle, and per unit time. BT is the ratio of the brilliance at the sample location relative to the source brilliance~\cite{andersen2018optimization}. BT under passive transport has a theoretical maximal value of 100\% due to Liouville's theorem~\cite{landau2013statistical} and thus is an informative metric for evaluating neutron transport efficiency. 

Figure~\ref{fig:BT}(a) shows the wavelength-resolved BT. The guide systems show the lowest BT at the short wavelength limit of 1.0~\AA, which gradually improves with increasing wavelengths until saturation. The BTs saturate around 4.8~\AA, 1.3~\AA, 1.7~\AA, 2.3~\AA~with the saturated values of 93\%, 95\%, 82\%, 92\%, respectively, for the Montel, straight-elliptical, curved-tapered, and curved-elliptical beamline. The Montel beamline shows the strongest wavelength-dependent BT with the lowest transport efficiency for neutrons with wavelengths shorter than $<3.0$~\AA, and gravity effects cause the drop of BT above 4.8~\AA. The straight-elliptical beamline delivers a simple, homogeneous beam profile with near fourfold symmetry in the x-y plane, and it has the highest neutron transport efficiency in most of the full wavelength range of 1.0--6.0~\AA. The curved-tapered beamline improves the thermal-neutron transport efficiency compared to the Montel mirrors but has the lowest cold-neutron transport efficiency among the four systems. The neutron transport efficiency of the curved-elliptical beamline is between the straight-elliptical and the curved-tapered beamline. 

Figure~\ref{fig:BT}(b)-(e) shows the spatial distribution with the desired beam divergence and the desired wavelength of 1.0--6.0~\AA. The Montel mirrors deliver a uniform beam within the spatial ROI and a low flux outside---a desired outcome. The straight-elliptical beamline has a homogeneous profile with a quasi-fourfold symmetry in the x-y plane, but there are more neutrons outside the ROI than the Montel beamline. The straight-elliptical beamline has the brightest beam, corresponding to the highest BT value. The curved-tapered and curved-elliptical beamlines lead to a more spatially extended distribution of neutrons outside ROI. The curved-elliptical beamline shows stripe patterns along the vertical directions, indicating neutron flux inhomogeneity in the horizontal plane.

\subsection{Flux}
\begin{figure}
\includegraphics[width=0.5\textwidth]{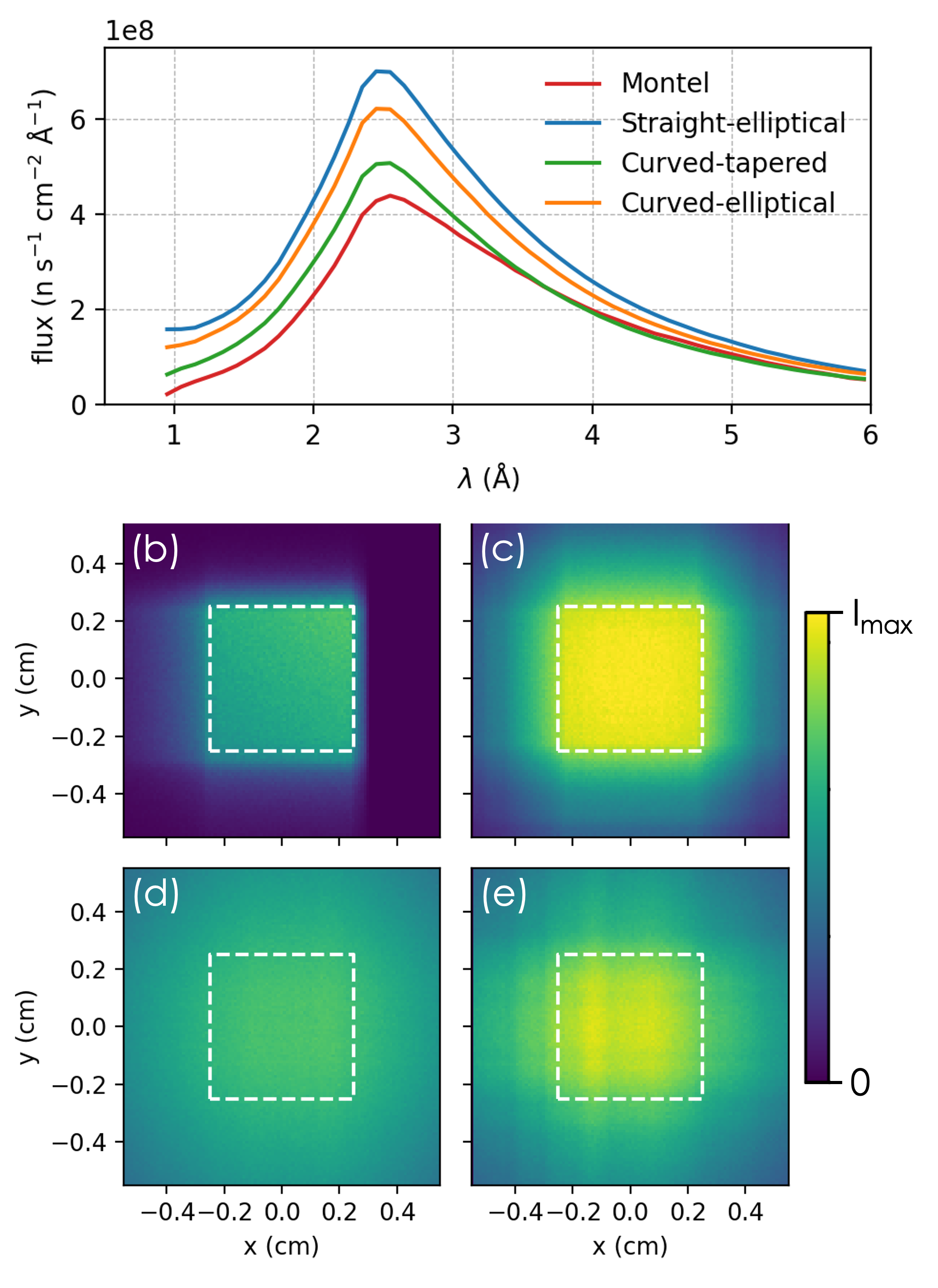}
\caption{\label{fig:flux} (a) Wavelength-resolved flux integrated over the spatial ROI. (b-e) The spatial distribution of the neutron flux with the wavelengths between 1.0~\AA~and 6.0~\AA~and integrated over all divergence. The plots use a bin size of $100\times100~\mu$m$^2$. From (b) to (e) are the results for the Montel, the straight-elliptical, the curved-tapered, and the curved-elliptical beamline, respectively. The white dashed boxes represent the spatial ROI.}
\end{figure}

While Fig.~\ref{fig:BT} includes only neutrons with the divergence ROI, Fig.~\ref{fig:flux} shows the flux and spatial distributions, integrated over all divergence, which reflects the flux at the sample location. The straight-elliptical, curved-tapered, and curved-elliptical beamlines show a flux peaked at  2.5~\AA~, which coincides with that from the source spectrum. This is consistent with the slowly varying BT curves above 2.0~\AA~for these beamlines, as seen in Fig.~\ref{fig:BT}(a). In contrast, the Montel beamline shows a flux peaked at 2.6~\AA, reflecting the strong wavelength-dependent neutron transport efficiency. The flux profiles are homogeneous within the spatial ROI for the straight-elliptical and the curved-tapered beamlines. However, the Montel beamline shows more neutrons in the top-right corner in the spatial ROI, and the curved-elliptical beamline shows more pronounced vertical strip patterns. The non-uniform flux distribution is easier to view in Fig.~\ref{fig:SMFlux}.

\subsection{\label{sec:acceptdia}Acceptance Diagram}
\begin{figure*}
\includegraphics[width=0.9\textwidth]{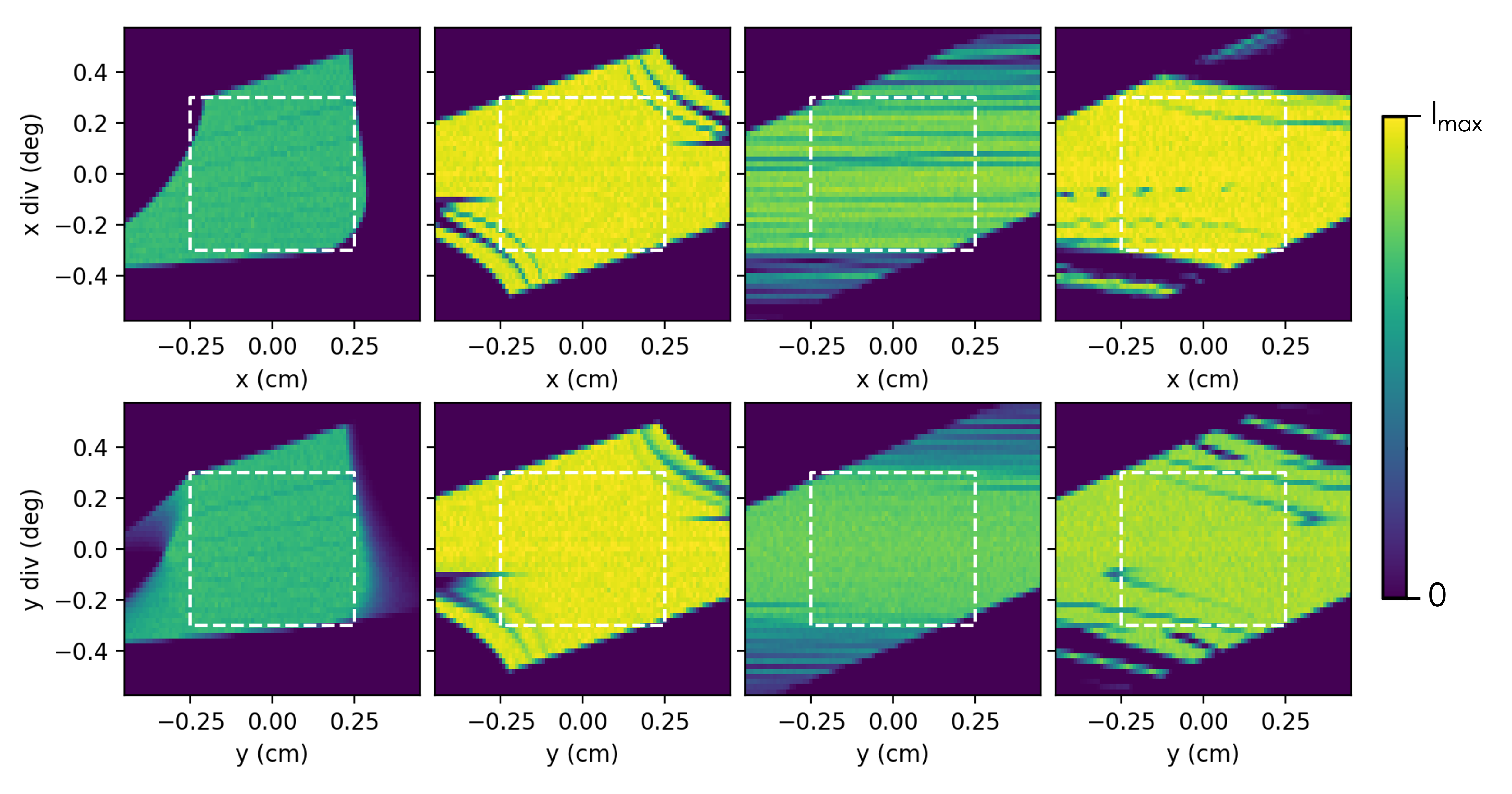}
\caption{\label{fig:PhaseSpace} The acceptance diagrams at the sample position. The top (bottom) rows are the acceptance diagrams along the horizontal (vertical) direction, with a bin size of $100~\mu$m $\times~0.02^\circ$. From the left to the right, the four columns show the results of the Montel, the straight-elliptical, the curved-tapered, and the curved-elliptical beamline, respectively. The white dashed boxes highlight the desired ROIs. The plots reflect the wavelength-integrated beam profiles over 1.0--6.0~\AA.}
\end{figure*}

\begin{figure}
\includegraphics[width=0.50\textwidth]{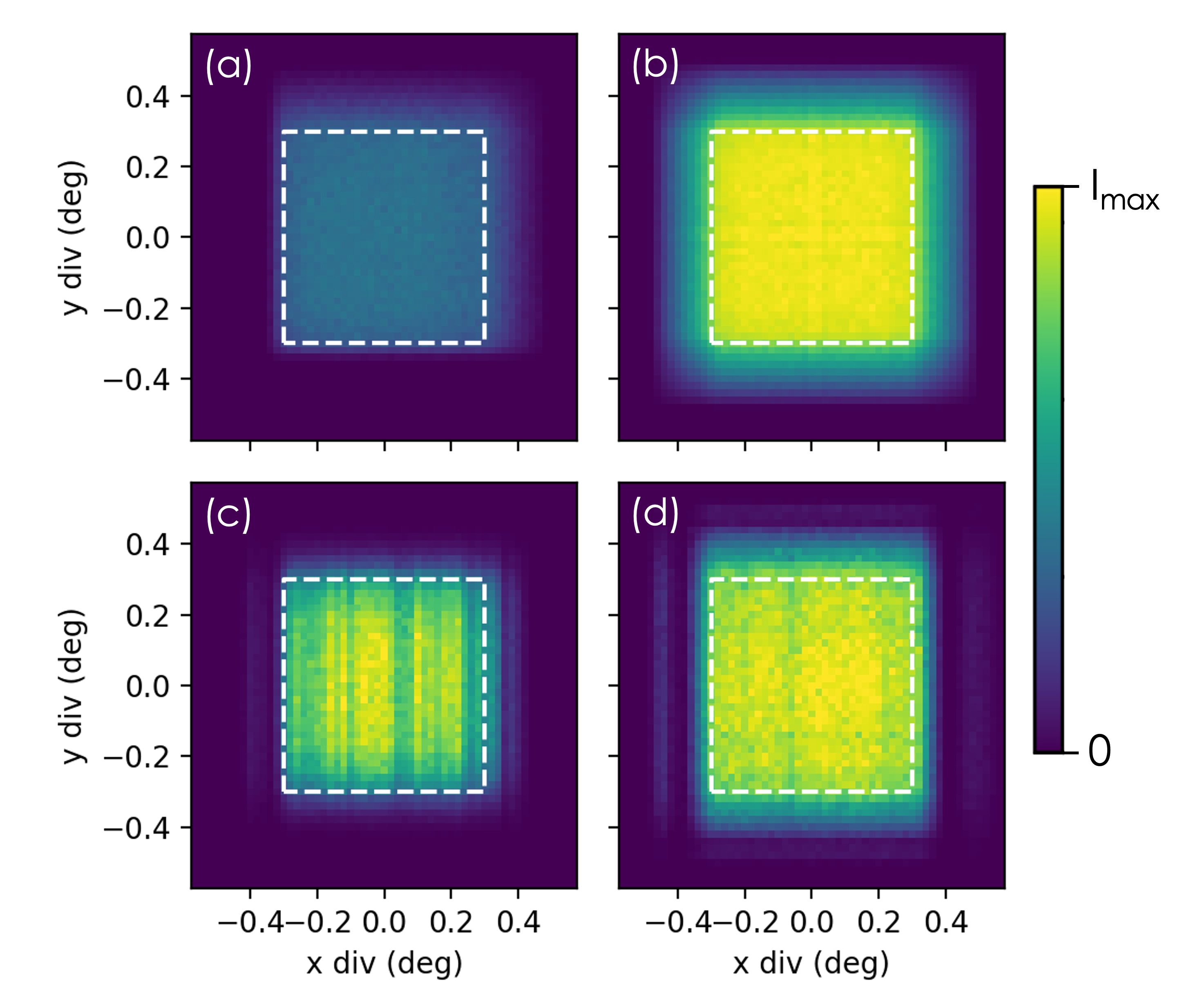}
\caption{\label{fig:DivMap} The angular distribution of the wavelength-integrated flux over 1.0--6.0~\AA~within the spatial ROI at the sample position. The bin size is $0.02^\circ \times 0.02^\circ$. From panel (a) to (d) are the results of the Montel, the straight-elliptical, the curved-tapered, and the curved-elliptical beamline, respectively.}
\end{figure}

\begin{figure*}
\includegraphics[width=0.85\textwidth]{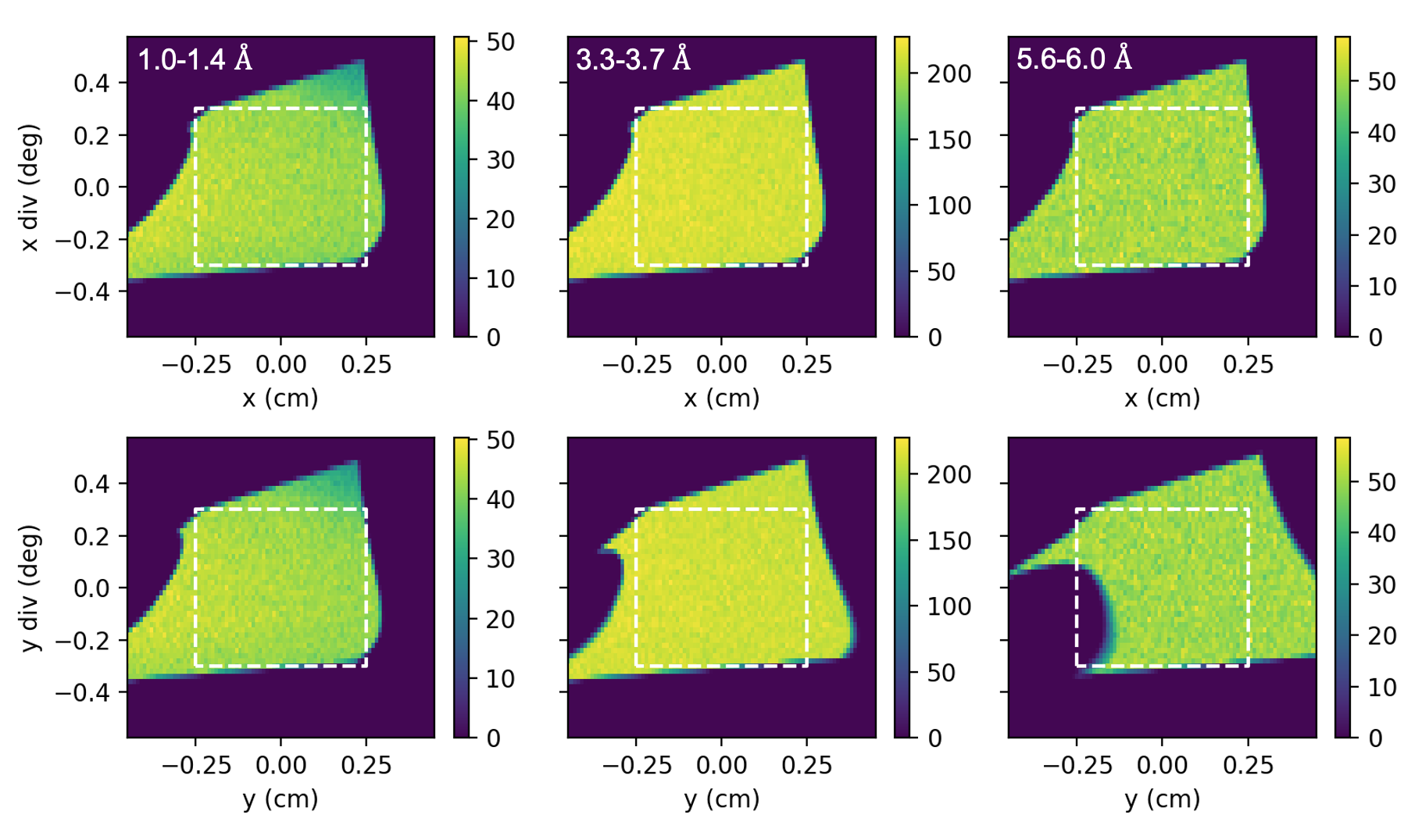}
\caption{\label{fig:WL_PhaseSpace_Montel} The wavelength-dependent acceptance diagrams of the Montel beamline, where the top and bottom rows are the results in the horizontal and the vertical direction, respectively.  From left to right, the results from a 0.4-\AA~wavelength band centered at 1.2~\AA, 3.5~\AA~and 5.8~\AA, respectively. Different color scales are used due to the wavelength-dependent spectrum.}
\end{figure*}

\begin{figure*}
\includegraphics[width=0.85\textwidth]{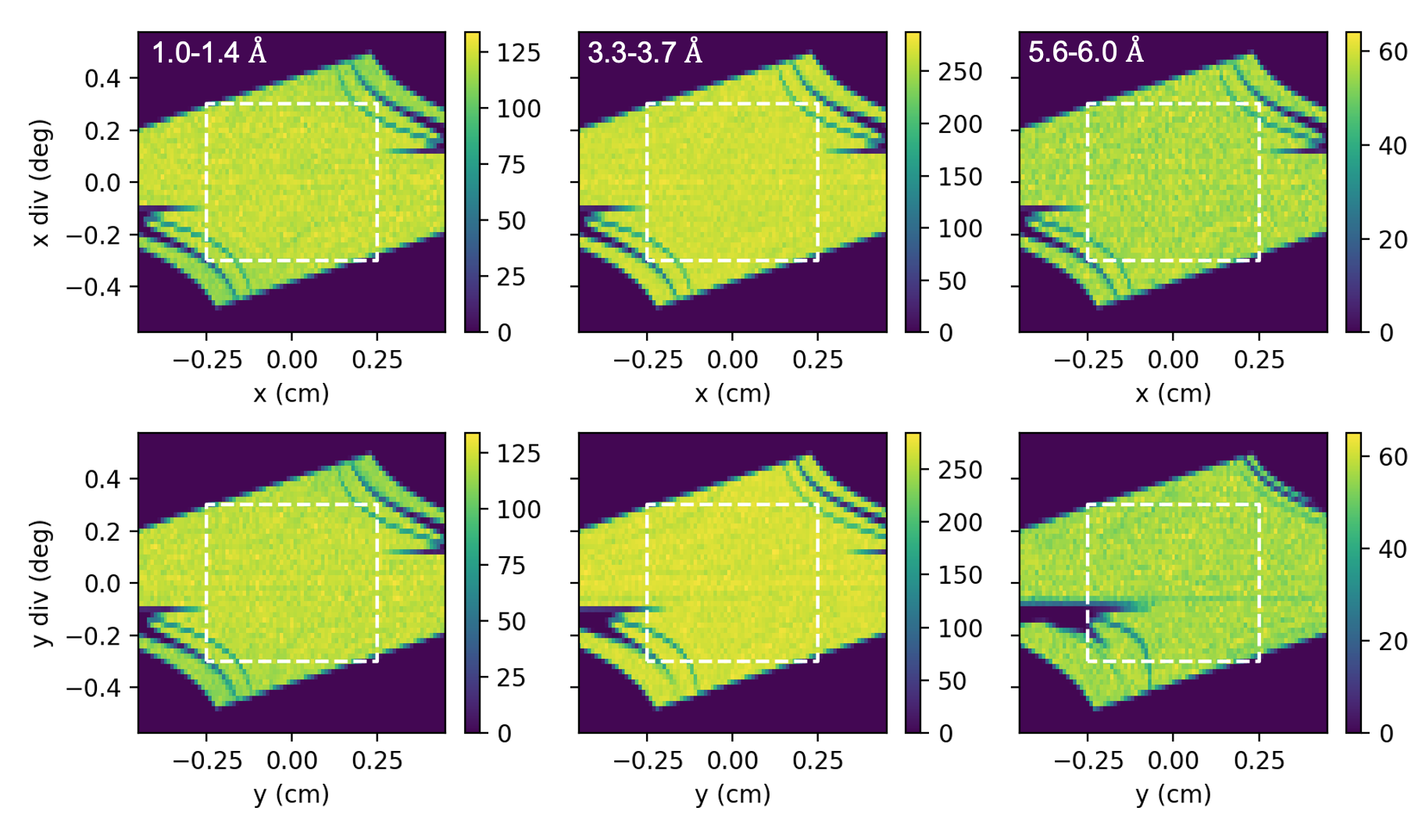}
\caption{\label{fig:WL_PhaseSpace_Straight} The wavelength-dependent acceptance diagrams of the straight-elliptical beamline, using the same specification as Fig.~\ref{fig:WL_PhaseSpace_Montel}.}
\end{figure*}

Figure~\ref{fig:PhaseSpace} shows the acceptance diagrams at the sample position. The white boxes present the ROI. The angular and spatial distributions of neutrons shall be homogeneous within the ROI, and the neutron density shall quickly diminish outside it.  

The Montel beamline shows a homogeneous distribution with the ROI. However, lacking the mirror symmetry in both the horizontal and vertical directions leads to more flux distributions with positive angles than with negative angles in both the horizontal and vertical directions, which is better visualized in Fig.~\ref{fig:DivMap}(a). These neutrons contribute to the increased flux in the top-right corner of the spatial ROI seen in Fig.~\ref{fig:flux}(b). There are small voids at the two corners of the phase space in the acceptance diagram, i.e., at [negative-position, positive-angle] and [positive-position, negative-angle], caused by the competing effect of removing unwanted neutrons outside the ROI.  

The acceptance diagram of the Montel beamline has a sharp boundary in the horizontal plane and a blurry boundary in the vertical direction. Note that the mirror geometries are identical in the horizontal and the vertical directions, suggesting that the difference originates from gravity effects that mostly affect the vertical component of neutron trajectories. This can be seen from the wavelength-resolved acceptance diagrams, as shown in Fig.~\ref{fig:WL_PhaseSpace_Montel}, because slower neutrons' trajectories are more sensitive to gravity effects. The acceptance diagrams are nearly wavelength-independent in the horizontal direction but change with wavelengths in the vertical direction. Therefore, including neutrons of a broad range of wavelengths leads to edge blurriness in the vertical acceptance diagram. Furthermore, voids appear at the long wavelength limit in the vertical acceptance diagram, decreasing BT above 4.8~\AA. The acceptance diagrams of other beamlines show much weaker gravity effects. Figure~\ref{fig:WL_PhaseSpace_Straight} shows the straight-elliptical beamline as an example. 

The straight-elliptical beamline acceptance diagrams along the horizontal and vertical directions show a quasi-inversion symmetry because of symmetric mirror configuration in both the horizontal and vertical directions and much weaker gravity effects. The acceptance diagrams are uniform within the ROI, only including minor voids. This indicates that the tapered, straight guide sections have effectively filled the gaps typically seen in guides only containing elliptical sections~\cite{weichselbaumer2015tailoring, ma2021performance}. 

The curved-tapered beamline's acceptance diagram shows high-frequency structures, which is not originated from imperfect reflectivity, as shown in Fig.~\ref{fig:SMCurved}. The horizontal angular distribution displays a flux modulation with a period of $\sim 0.043(2)^\circ$ (see Fig.~\ref{fig:SMCurved_Divfreq} for details). There are no such high-frequency structures in the vertical direction within the ROI and turning off gravity effects barely changes the distribution, suggesting that the inhomogeneity is mainly related to the curved guide section, which is known to cause phase-space inhomogeneity~\cite{mildner1990acceptance, mildner2008curved}.  However, such a high-frequency inhomogeneity will not affect diffraction data quality for most cases, considering the convolution effect from a typical crystal mosaicity of a few tenths of a degree. At the same time, the waviness of real guide systems will smooth the angular distribution, making the high-frequency pattern less observable, as demonstrated in Fig.~\ref{fig:SMCurved}.

The acceptance diagrams of the curved-elliptical beamline are similar to those of the straight-elliptical beamline, but the profiles become inhomogeneous with the non-straight guide section. Furthermore, the inversion symmetry in the horizontal direction seen in the straight-elliptical beamline is broken.  

\section{Post Optimization Analysis}\label{sec:perf}
\subsection{Supermirror Reflections}
\begin{figure}
\includegraphics[width=0.45\textwidth]{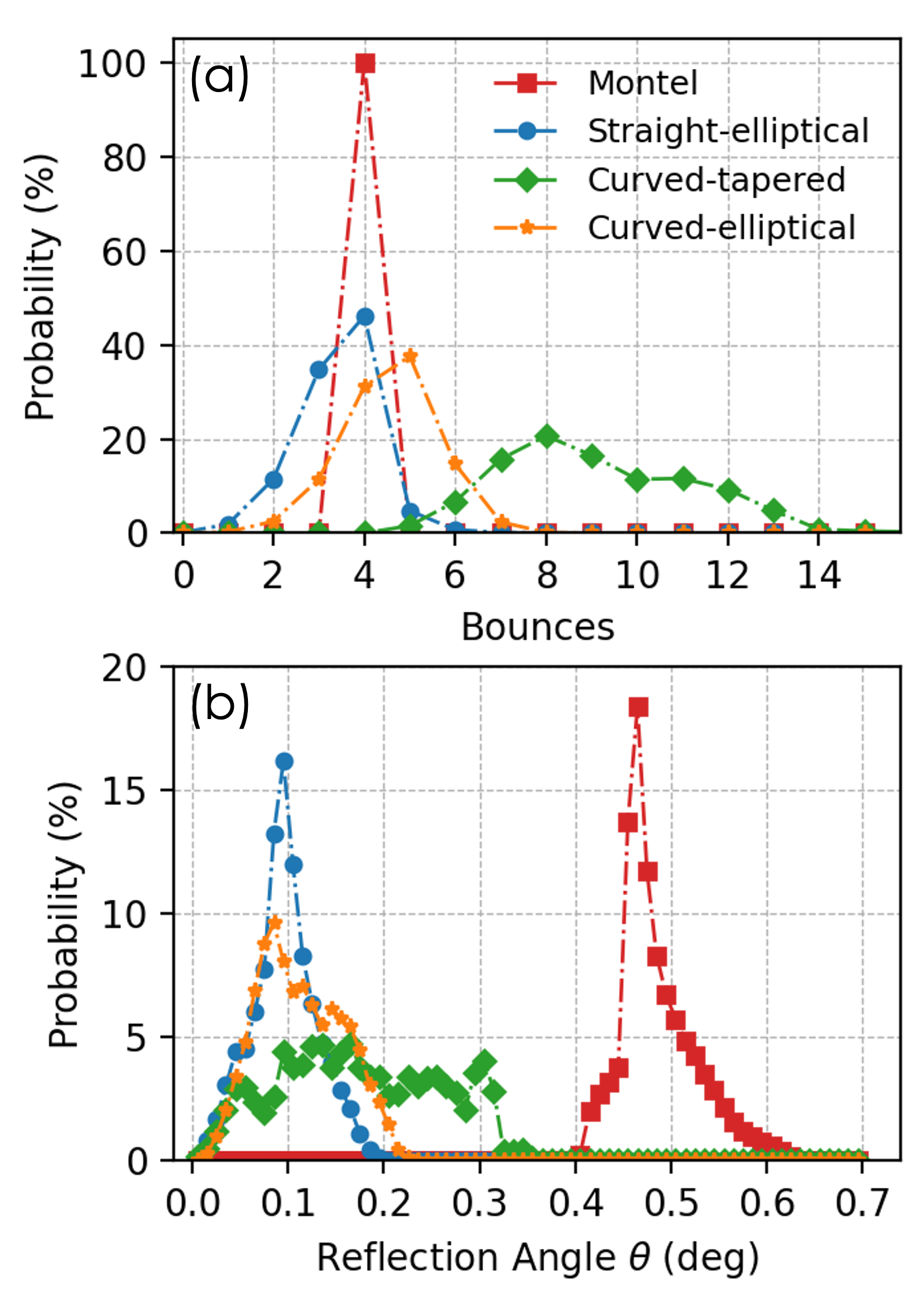}
\caption{\label{fig:Analysis_Refl} The statistics of (a) the bounces and (b) the supermirror reflection angles for neutron rays transported within the desired phase space ROI for the four guides. The angular bin size is 0.01$^{\circ}$ in panel (b). Only neutrons within 1.0--6.0~\AA~and delivered into the phase space ROI are considered. }
\end{figure}

We have analyzed the details of the supermirror reflections to understand the distinct neutron transport performance of the four guides. Figure~\ref{fig:Analysis_Refl} shows the statistics of the neutron bounces and reflection angles. Neutrons have a non-unity possibility of being specularly reflected when they hit the guide surface. As shown in Fig.~\ref{fig:Refl}, the reflectivity is a function of the momentum transfer $Q$ on specular reflection. $Q = 4 \pi sin(\theta_f)/\lambda$, depending on the reflection angle $\theta_f$ and the wavelength $\lambda$. A shorter-$\lambda$ reflection along a fixed neutron trajectory is associated with a higher $Q$. Therefore, shorter-$\lambda$ neutrons will typically have lower possibilities of being confined within the guide and transported over a long distance, leading to a lower BT.  As shown in Fig.~\ref{fig:Refl}, supermirrors with m = 6 support high reflectivities up to ~0.13~\AA$^{-1}$, corresponding to a reflection angle of 0.59$^\circ$ for 1~\AA~neutrons. Therefore, based on the reflection angle distributions (Fig.~\ref{fig:Analysis_Refl}(b),  the supermirror coating value (m = 6) used in simulations is sufficient to transport neutrons in the wavelength range of 1-6~\AA, except for the Montel beamline, which will be slightly impacted in the short-wavelength limit.

The Montel beamline is a good case to illustrate the effects of imperfect reflectivity. We have tracked all reflection events for neutrons within the ROI. These neutrons are precisely bounced four times before reaching the sample position, but $\theta_f$ varies significantly, with an average value of 0.49$^\circ$. We can estimate the average loss per bounce by converting the average $\theta_f$ into $Q$ and then estimate the mean reflectivity. This translates to an average of 28\% loss per bounce for 1.0-\AA~neutrons, and there will be less than a 27\% probability remaining in the guide after four bounces. In contrast, the average $Q$ for 5.0-\AA~neutrons is close to $q_{C}^{Ni}$. Therefore, there is only a small percentage of loss after four bounces. This explains the strong $\lambda$-dependent BT for the Montel beamline.

The average bounces are 3.4, 4.6, and 9.2, and the average supermirror $\theta_f$ are 0.10$^\circ$, 0.11$^\circ$, and 0.17$^\circ$ for the straight-elliptical, curved-elliptical, and curved-tapered beamlines, respectively. The average reflection angles are much smaller than that in the Montel beamline, which leads to smaller $Q$ and fewer neutron losses per bounce. Figure~\ref{fig:Refl} shows the average $Q$ from all four guide systems using three representative wavelengths, 1.0~\AA, 2.5~\AA~and 5.0~\AA. The straight-elliptical beamline has the smallest average $\theta_f$ and the fewest bounces. Therefore, supermirror reflections from the optimized straight-elliptical beamline will have the least flux loss effect, which is the main factor for the straight-elliptical beamline's superior transport efficiency. 

\subsection{Phase Space Homogeneity}
\begin{figure}
\includegraphics[width=0.44\textwidth]{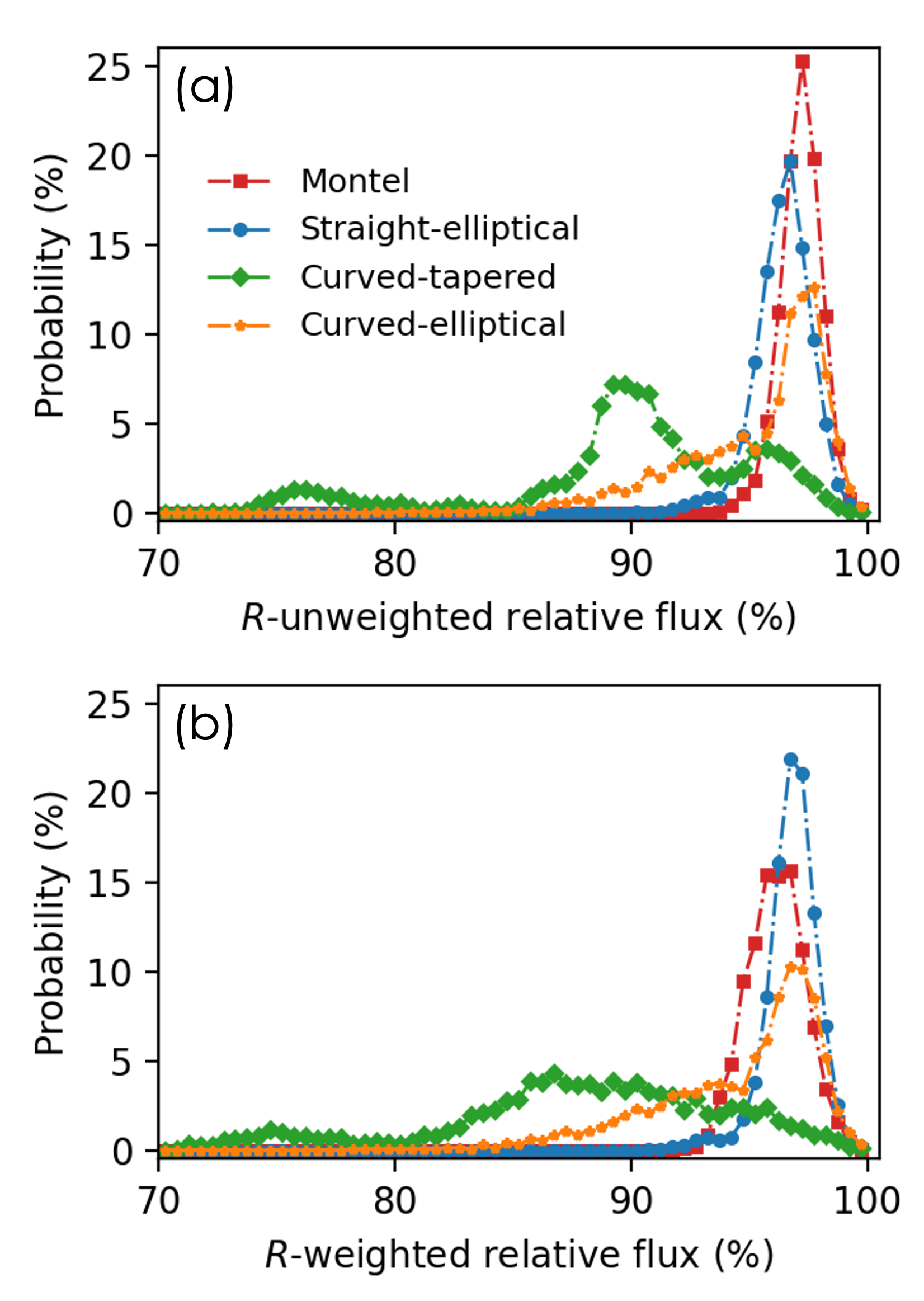}
\caption{\label{fig:Analysis_PhaseSpace} Histogram of the flux distribution within the central 4D phase space of $4\times4$~mm$^{2}$ area and a $\pm 0.25^\circ$ beam divergence in both the horizontal and vertical directions. The phase space has been evenly divided into 4096 subspaces for statistical analysis. The neutron flux is integrated over a 1.0--6.0 \AA~wavelength band, assuming (a) perfect reflectivity ($R$-unweighted) and (b) imperfect reflectivity ($R$-weighted). Flux values are normalized to the highest value from the 4096 subspaces and then binned into a histogram with a bin size of 0.5\% of the highest flux value.}
\end{figure}

Phase space homogeneity at the sample location also affects BT, provided that the neutron source has a uniform surface brilliance within the view area.  Liouville's theorem holds for any subspace of the desired phase space at the sample location. Therefore, if there is significant inhomogeneity within the desired phase space, some subspaces must have low BT values; thus, the average BT over the entire desired phase space will be low. 

Figure~\ref{fig:Analysis_PhaseSpace} shows the flux distribution of the phase space of the four guide systems by statistical analysis of the number of neutrons in fine-binned phase space. Both the scenarios of perfect and imperfect reflectivity have been studied. Assuming perfect reflectivity, the fluxes of the Montel and straight-elliptical beamlines show a single-modal distribution with a relative standard deviation (rSTD) of 0.820(4)\% and 1.064(5)\%, respectively, where rSTD is the ratio between the flux deviation and the peak flux extracted from a Gaussian curve fit. In contrast, the curved-tapered and curved-elliptical beamlines show multi-modal and broader distributions. 

Since Monte Carlo ray tracing is a stochastic process, there is a flux distribution width due to simulations of finite rays. This study's ROI includes a total neutron ray count of $6.7~\times 10^{7}$ for analysis. If these rays were randomly distributed over 4096 bins, it would lead to an rSTD of 0.782\%. This is very close to the Montel beamline case assuming perfect reflectivity, which can be attributed to the fact that each neutron delivered into the ROI is bounced exactly four times for the Montel beamline, once per mirror surface. However, the Montel beamline shows a large variation of $\theta_f$; therefore, neutron rays reaching different regions of the phase space have different attenuation effects from supermirror reflection.  As a result, the flux distribution of the Montel beamline becomes broader with a rSTD of 1.289(8)\% after considering the imperfect reflectivity. 

After considering the imperfect reflectivity, the straight-elliptical beamline shows a reduced rSTD of 0.895(4)\%. This is likely because the optimization metric considers the phase space homogeneity based on imperfect reflectivity. Overall, the optimized straight-elliptical beamline's phase space becomes the most homogeneous one among the four beamlines. 

The curved-elliptical beamline shows a broader range of flux distribution than the straight-elliptical beamline. The main difference between the geometry of the two beamlines is the tapered guide section, which suggests that the non-straight guide section increases the phase space inhomogeneity.

The curved-tapered beamline shows the broadest, multi-modal flux distribution before considering imperfect reflectivity. Therefore, such non-uniformity is mainly from the guide geometry. After taking account of the imperfect reflectivity, the flux distribution also shows the most significant broadening among the four beamlines, which correlates with that the curved-tapered beamline shows the broadest distribution of the bounces and $\theta_f$, as seen in Fig.~\ref{fig:Analysis_Refl}.

\subsection{Additional discussion}
The straight-elliptical beamline performs best, showing the highest transport efficiency for cold/thermal neutrons and uniform phase space. Using a pair of half-ellipse guides can reduce the aberration effect from the finite source size~\cite{maier1963use}, which also applies to the curved-elliptical and Montel beamlines. 

The curved-elliptical beamline is modified from the straight-elliptical beamline. The kinked angle between the two major axes of the two half-ellipse guides has been optimized for neutron transport performance. The non-straight guide section eliminates the direct line of sight from the moderator; however, it increases the bounces and the reflection angles, deteriorates the phase-space homogeneity, and leads to more neutrons outside the spatial ROI. 

The Montel beamline provides a homogeneous phase space and shows the best performance of confining neutrons within the spatial ROI. However, it only uses two reflection surfaces, resulting in higher average reflection angles, which limits its ability to transport thermal neutrons. At the same time, such a large Montel guide optic has not been extensively tested, presenting a high risk. The acceptance diagram's edge blurriness is only observed for the Montel beamline, indicating that gravity effects do not significantly affect the phase space for other options in the wavelength range of interest. Therefore, a guide system's closed or non-closed cross-section affects gravity effects on neutron transport. 

The curved-tapered beamline shows the lowest cold neutron transport efficiency. We are considering a long sample-to-source distance with guides of small cross-sections. Therefore, the direct line of sight from the moderator is easily avoided. It is known that a curved guide will cause significant phase space inhomogeneity~\cite{mildner1990acceptance}. The linearly taped convergent guide after the bender can focus the beam and improve the spatial homogeneity~\cite{mildner2008curved}; however, it does not solve the angular distribution inhomogeneity~\cite{schanzer2004advanced}; instead, it leads to the observed high-frequency stripe patterns. As shown in Fig,~\ref{fig:DivMap}, only the curved-tapered beamline shows such a notable stripe pattern in the angular flux distribution. Such high-frequency non-uniform angular distribution has little effect on scattering patterns but will reduce the brilliance at the sample location.  

\section{Summary}
We have compared four general guide concepts using the STS single-crystal diffractometer concept. We found that the straight-elliptical beamline with a pair of half-ellipse sections has the best neutron transport efficiency because neutrons experience the fewest bounces and the smallest reflection angles, reducing the beam loss. We have also found that both guide geometries and imperfect mirror reflections affect phase-space homogeneity. The straight-elliptical beamline has a simple geometry with a narrow distribution of bounces and reflection angles, so that it delivers the most uniform phase space within the region of interest among the four guides, making it the best choice for Pioneer. The results can be applied to general neutron instruments facing a compact, high-brilliance moderator, using cold and thermal neutrons, requiring a moderate and uniform phase-space volume for experiments. 

\section{Supplementary Material}
The supplementary materials include four figures.  Figure~\ref{fig:workflow} explains the workflow used for final optimizations (Sec.~\ref{sec:method_opt}). Figure~\ref{fig:SMFlux} shows the spatial inhomogeneity for all optimized guides (Fig.~\ref{fig:flux}). Figure~\ref{fig:SMCurved} compares the acceptance diagrams of the curved-tapered beamlines under different assumptions about gravity effects and the supermirrors' waviness and reflectivity (Sec.~\ref{sec:acceptdia}). Figure~\ref{fig:SMCurved_Divfreq} shows the high-resolution angular distribution of the curved-tapered beamline along the horizontal and vertical directions (Sec.~\ref{sec:acceptdia}). 
 
\begin{acknowledgments}
The author thanks Peter Torres, John Ankner,  Leighton Coates, and Kenneth Herwig for discussions. This research used resources from the Spallation Neutron Source Second Target Station Project at Oak Ridge National Laboratory (ORNL). ORNL is managed by UT-Battelle LLC for DOE's Office of Science, the single largest supporter of basic research in the physical sciences in the United States. 

\end{acknowledgments}

\section*{Data Availability Statement}
The data supporting this study's findings are available from the corresponding author upon reasonable request.

\bibliography{Pioneer}

\pagebreak
\renewcommand{\thefigure}{S\arabic{figure}}
\setcounter{figure}{0}

\begin{CJK*}{GB}{} 

\title[]{Supplementary Materials: General Guide Concepts for Compact, High-Brilliance Neutron Moderators}
\author{Yaohua Liu}
\email[The author to whom correspondence may be addressed: ]{liuyh@ornl.gov}
\affiliation{Second Target Station, Oak Ridge National Laboratory, Oak Ridge, Tennessee 37831, USA}%
\date{\today}
\maketitle
\end{CJK*}

\begin{figure*}
\includegraphics[width=1.0\textwidth]{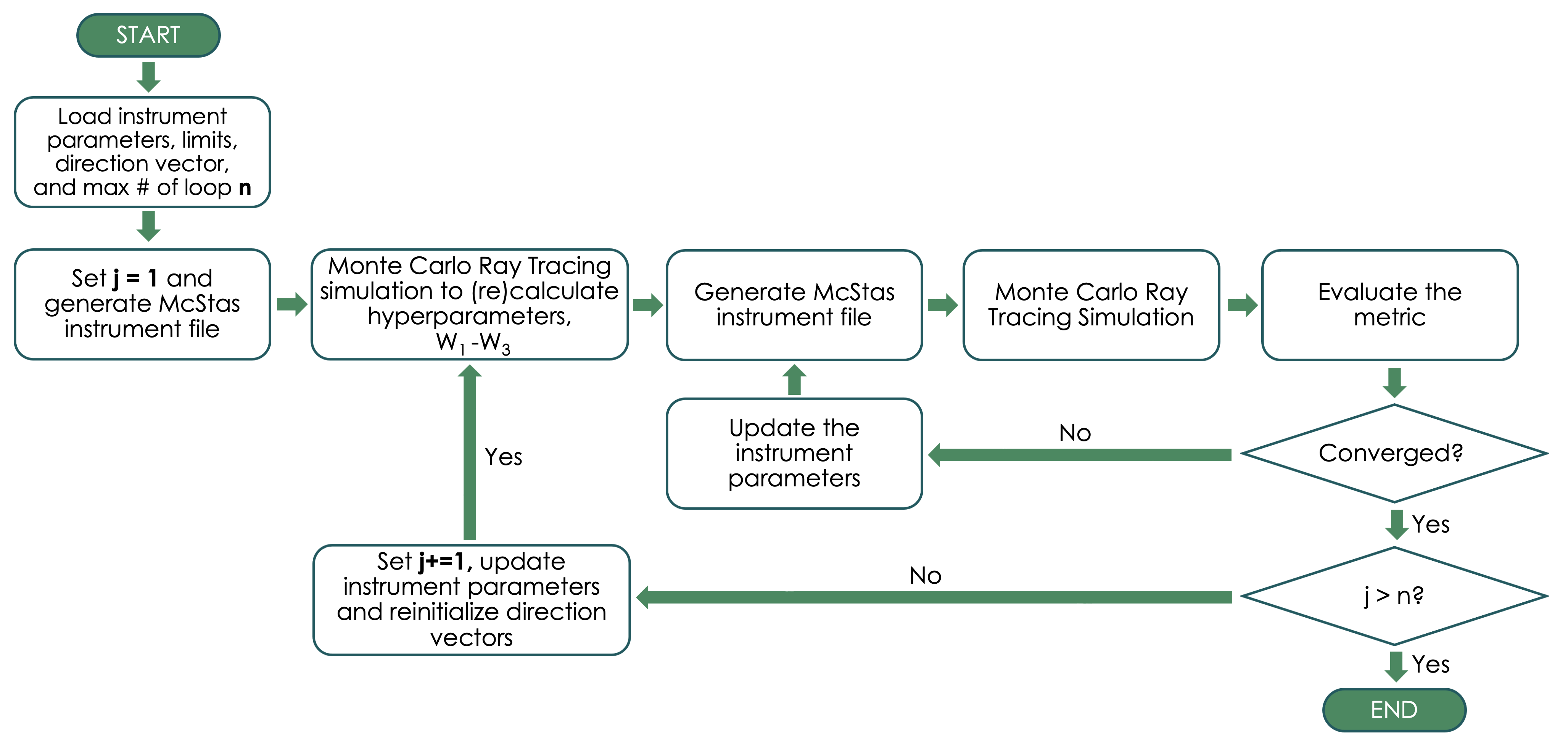}
\caption{\label{fig:workflow} The workflow used for final optimizations employs a double-loop structure. The outer loop updates hyperparameters and resets the direction vector with a typical n value ranging from 4 to 10, while the inner loop optimizes the instrument parameters. There are five items in the metric: (1) $I_{roi}$,  the integrated flux within the desired 4D phase space volume; (2)  $w_{0} \times I_{outside-div-roi}$ and  $w_{1} \times I_{outside-spatial-roi}$ are the weighted flux outside the angular and spatial ROIs; (4)  $w_{2} \times \delta I_{div-roi}$ is the weighted standard deviation of the angular distribution within the spatial ROI, and (5)  $w_{3} \times \delta I_{spatial-roi}$  is the weighted standard deviation of the spatial distribution within the spatial ROI. The value of $w_0$ is empirically chosen, with a typical value of 0.1. The values of $w_1$--$w_3$ are calculated from the Monte Carlo ray-tracing simulation result at the beginning of each major loop and kept fixed during the minor-loop optimization. The hyperparameters are chosen so that each weighted $w_1$ to $w_3$ term equals a pre-selected fraction (typically 10\% to 30\%) of $I_{roi}$.}
\end{figure*}

\begin{figure*}
\includegraphics[width=0.5\textwidth]{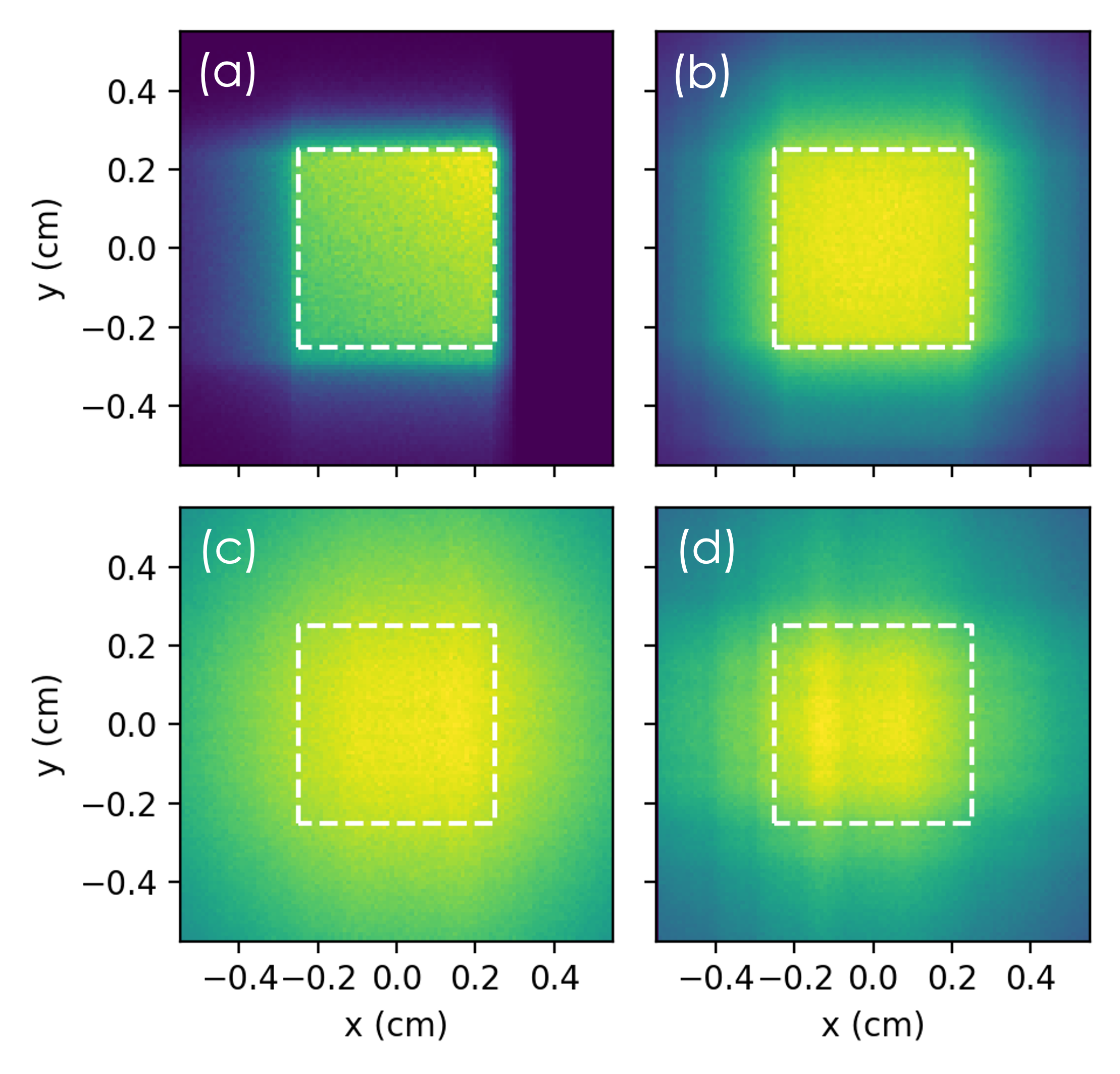}
\caption{\label{fig:SMFlux} The spatial flux distributions at the sample position of (a) the Montel, (b) the straight, (c) the curved, and (d) the kinked beamline, respectively. Figure~\ref{fig:flux} in the main text uses a common color scale to highlight the different flux levels among the four beamlines. Here, different color scales are used to emphasize the inhomogeneity. The dashed white boxes represent the ROI. The plots use a bin size of $100\times100~\mu$m$^2$.}
\end{figure*}

\begin{figure*}
\includegraphics[width=0.8\textwidth]{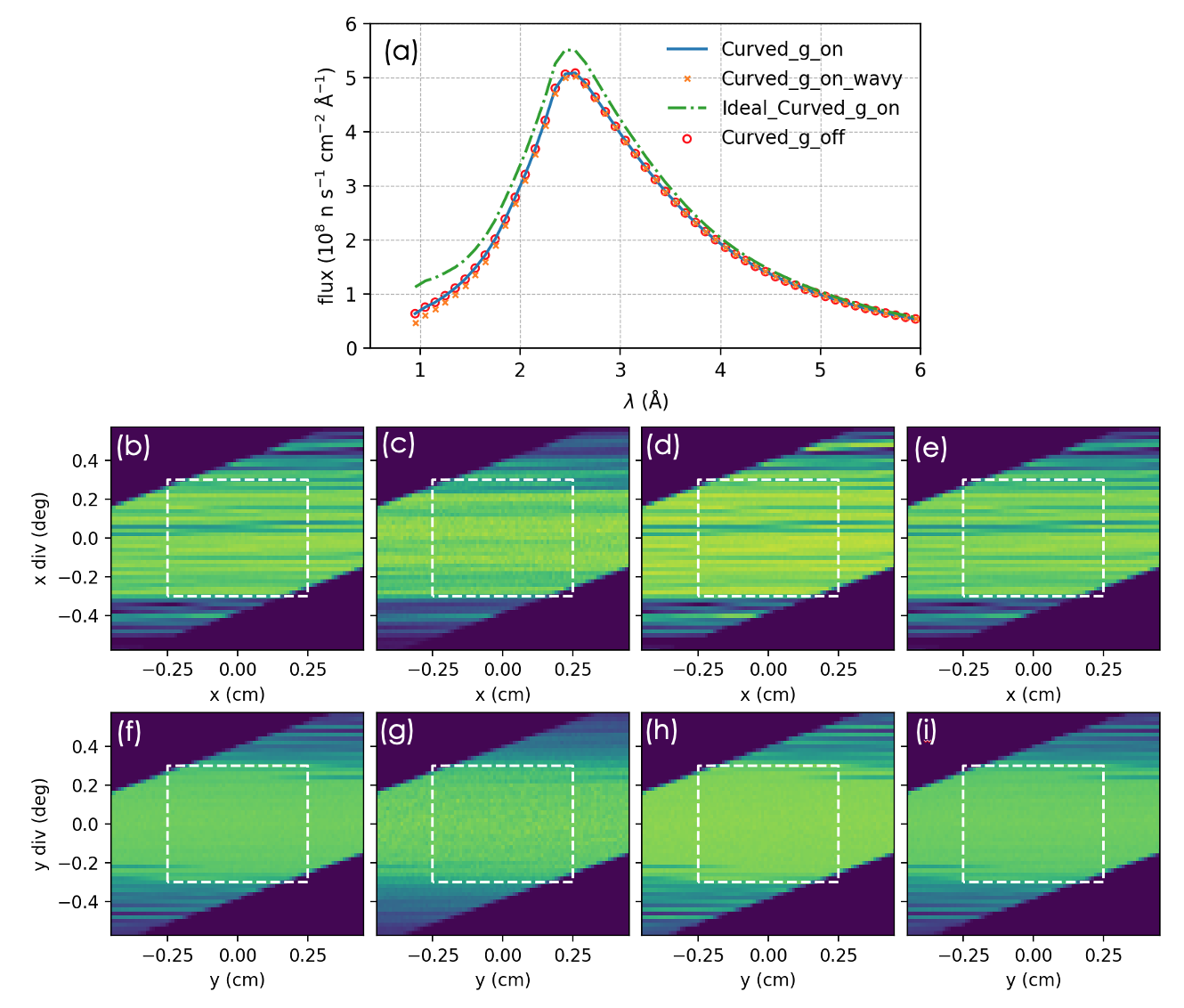}
\caption{\label{fig:SMCurved} (a) The spectra within the spatial ROI at the sample location of the curved beamline. The suffixes $g\_on$ and $g\_off$ indicate whether the simulations consider or ignore gravity effects. The prefix $ideal$ indicates that ideal supermirrors are assumed, ignoring imperfect reflectivity effects. The suffix $wavy$ indicates that the waviness effects of the supermirror coating are considered with a waviness value of $1\times10^{-4}$~ rad. (b)-(e) the acceptance diagrams in the horizontal direction, and (f)-(i) the acceptance diagrams in the vertical direction. (b, f) correspond to Curved$\_$g$\_$on; (c, g) correspond to Curved$\_$g$\_$on$\_$wavy; (d, h) correspond to ideal$\_$Curved$\_$g$\_$on; and (e, i) correspond to Curved$\_$g$\_$off. The acceptance diagrams use a bin size of $100~\mu$m $\times~0.02^\circ$.  }

\end{figure*}

\begin{figure*}
\includegraphics[width=0.8\textwidth]{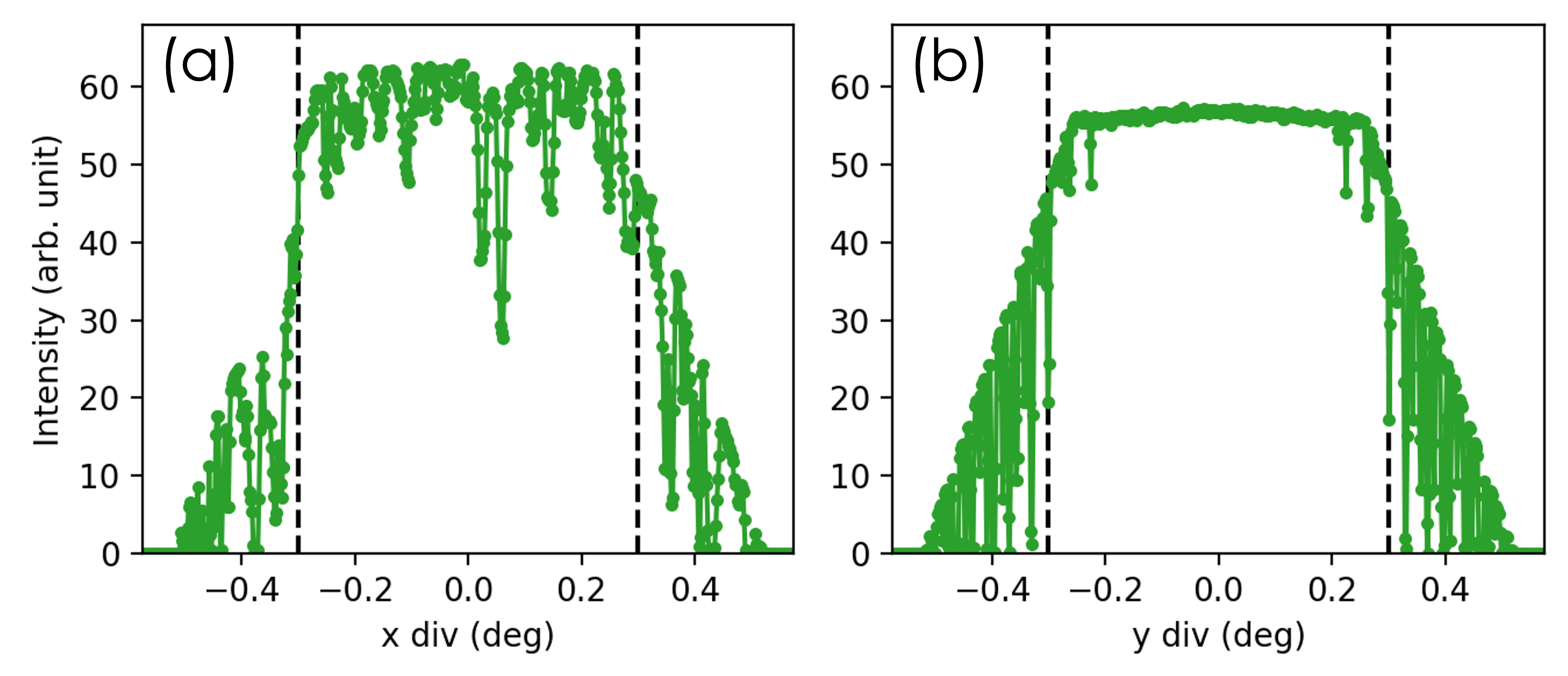}
\caption{\label{fig:SMCurved_Divfreq} The angular flux distribution within the spatial ROI of the curved beamline in (a) the horizontal  (x) direction and (b) the vertical (y) direction, with a bin size of 0.002$^\circ$. The curved$\_$g$\_$on case is used for analysis. The black dashed lines denote the angular ROI. The dominant-frequency component of the angular flux distribution within the ROI has a period of 0.043(2)$^{\circ}$ in the horizontal direction.}
\end{figure*}

\end{document}